\documentclass[twocolumn,showpacs,showkeys,preprintnumbers,amsmath,amssymb]{revtex4}
 \usepackage{dcolumn}% Align table columns on decimal point
 \usepackage{bm}% bold math
\usepackage{graphicx}
 
\begin{document}

 \title{Superradiant instability of D-dimensional
 Reissner-Nordstr\"{o}m black hole mirror system}

 \author{Ran Li}

 \thanks{Corresponding author. Electronic mail: 021149@htu.cn}

 \author{Junkun Zhao}

 \author{Yanming Zhang}

 \affiliation{Department of Physics,
 Henan Normal University, Xinxiang 453007, China}

 \begin{abstract}

 We analytically study the superradiant
 instability of charged massless scalar field in the background of
 D-dimensional Reissner-Nordstr\"{o}m (RN) black hole caused by mirror-like boundary condition.
 By using the asymptotic matching method to solve the Klein-Gordon equation
 that govern the dynamics of scalar field, we have derived the
 expressions of complex parts of boxed quasinormal frequencies,
 and shown they are positive in the regime of superradiance.
 This indicates the charged scalar field is unstable in
 D-dimensional Reissner-Nordstr\"{o}m (RN) black hole surrounded by mirror.
 However, the numerical work to calculate the boxed quasinormal frequencies
 in this system is still required in the future.

 \end{abstract}



 \maketitle

 Press and Teukolsky \cite{press} proposed to built \emph{black hole bomb}
 in terms of the classical superradiance phenomenon
 \cite{zeldovich,bardeen,misner,starobinsky} in 70s last century.
 When an impinging bosonic wave with 
 frequency satisfying superradiant condition is scattered by 
 event horizon of rotating black hole,
 the scattered wave will be amplified.
 If one places a reflecting mirror outside of the black hole, the
 amplified wave will be reflected into the black hole once again.
 So it is obvious that the bosonic wave will be bounced back and forth
 between event horizon and mirror.
 Meanwhile, the energy of the wave can become sufficiently large
 in this black hole mirror system until the reflecting mirror is destroyed.

 Recent studies of black hole bomb mechanism was initiated by Cardoso et. al. in \cite{cardoso2004bomb}. See also the Refs.\cite{Rosa,Lee,leejhep,jgrosa,hod2013prd,hodbhb}
 for the recent studies on this topic.
 Black hole bomb mechanism can also be realized in the following two cases.
 In the first case, the mass term of bosonic field plays the role of reflecting
 mirror, for example, the systems investigated in \cite{kerrunstable,detweiler,strafuss,dolan,
 Hod,hodPLB2012,konoplyaPLB,DiasPRD2006,zhangw,dolanprd2013}.
 The second case is to study bosonic field perturbation in black hole background with
 Dirichlet boundary condition at the asymptotic infinity.
 These background spacetimes include the rotating black holes in AdS spacetime \cite{cardoso2004ads,cardoso2006prd,KKZ,aliev,uchikata,rlplb,zhang},
 in G\"{o}del universe \cite{knopolya,rlepjc}, and in the linear dilaton
 background \cite{clement,randilaton}. In all these cases, superradiance will trigger the instabilities of rotating black holes plus bosonic field perturbation.

 For a charged scalar wave in the background of spherical symmetric charged black hole,
 the wave scattered by event horizon will also undergo superradiant process if the frequency of this impinging wave satisfying superradiant condition \cite{bekenstein}.
 However, it was proved by Hod in \cite{hodrnplb2012,hodrnplb2013} that,
 for four dimensional RN black holes,
 the existence of a trapping potential well outside black hole
 and superradiant amplification of trapped modes cannot be satisfied simultaneously.
 This means that the four dimensional RN black holes are stable under the perturbations of massive charged scalar fields. Soon after, Degollado et. al. \cite{Degolladoprd,Degollado} found that the same system can be made unstable by adding a mirror-like boundary condition like the case of the Kerr black hole.
 In \cite{liprd}, we also shown that the mass term of scalar field in charged stringy black hole is never able to generate a potential well outside the event horizon
 to trap superradiant modes. This is to say that
 the charged stringy black hole is stable against massive charged
 scalar perturbation. In \cite{ranliprogress,ranliplb}, we have further studied the superradiant instability of massless scalar field in the background of charged stringy black hole
 due to a mirror-like boundary condition. However, it is still interesting to
 study the high dimensional cases.

 More recently, Wang and Herdeiro \cite{mengjieprd} studied
 the superradiant instability of a charged scalar field in D-dimensional
 RN Anti-de Sitter black hole. According to the mechanism of black hole bomb,
 this work can also be generalized to study the superradiant instabilities of
 D-dimensional RN black hole caused by the mass of scalar field and 
 mirror's boundary condition. In \cite{zhangw}, the authors have
 discussed superradiant instability of extremal brane-world RN black hole
 against the charged massive scalar perturbation. 
 Scalar field quasinormal modes in the dyadosphere spacetime of charged black hole are studied by using the third-order WKB approximation in \cite{chunyan}.
 Quasinormal Modes of Phantom Scalar Perturbation in RN Black Hole
 was studied in \cite{Pan}. Hawking Radiation of charged GHS black hole 
 was also recently studied in \cite{GHSH}.

 In this paper, we will further study the system composed by
 D-dimensional RN black hole, charged massless scalar field, and 
 reflecting mirror outside of the black hole. By using the asymptotic matching method to solve the Klein-Gordon equation that govern the dynamics of scalar field, we will derive the
 expressions of the complex parts of boxed quasinormal frequencies,
 and show the corresponding instability caused by the reflecting mirror.

 The D-dimensional RN black hole \cite{MP} is described by the line element
 \begin{eqnarray}
 &&ds^2=-f(r)dt^2+\frac{dr^2}{f(r)}+r^2d\Omega_n^2\;,\\
 &&f(r)=1-\frac{\mu}{r^{n-1}}+\frac{q^2}{r^{2(n-1)}}\;,
 \end{eqnarray}
 where the parameter $n=D-2$ is introduced to parameterize the dimension of spacetime
 for later convenience. $d\Omega_n^2$ denotes the line element of the n-dimensional unit sphere,
 \begin{eqnarray}
 d\Omega_n^2=d\theta_{n-1}^2+\sin^2\theta_{n-1}(d\theta_{n-2}^2+\sin^2\theta_{n-2}
 (\cdots&&\nonumber\\
 +\sin^2\theta_2(d\theta_1^2+\sin^2\theta_1 d\phi^2)\cdots))\;,&&
 \end{eqnarray}
 where the ranges of azimuthal coordinates are given by $\phi\in[0,2\pi]$ and $\theta_i\in[0,\pi] (i=1,\cdots,n-1)$. The Maxwell gauge potential is given by
 \begin{eqnarray}
 A=-\sqrt{\frac{n}{2(n-1)}}\frac{q}{r^{n-1}}dt\;.
 \end{eqnarray}
 The event horizon $r_+$ is determined as the largest root of the equation $f(r)=0$.
 The parameters $\mu$ and $q$ are related to the mass $M$ and the charge $Q$ of the black hole towards the following relations
 \begin{eqnarray}
 &&\mu=\frac{4M}{nV_n}\;,\nonumber\\
 &&q^2=\frac{2Q^2}{n(n-1)}\;,
 \end{eqnarray}
 where $V_n=2\pi^{(n+1)/2}/\Gamma((n+1)/2)$ is the volume of the
 $n-$dimensional sphere.

 We consider the charged massless scalar field perturbation in the background of
 D-dimensional RN black hole. The dynamics is then governed by the corresponding
  charged Klein-Gordon equation
 \begin{eqnarray}
 \frac{1}{\sqrt{-g}}D_\mu\left[\sqrt{-g}g^{\mu\nu}D_\nu\right]\Psi=0\;,
 \end{eqnarray}
 where $D_\mu=\partial_\mu-ieA_\mu$ with $e$ being the charge of the scalar field.
 The equation can be decomposed by the separation of variables for $\Psi$ as follows
 \begin{eqnarray}
 \Psi(t,r,\theta_i,\phi)=e^{-i\omega t} R(r) Y_{l,n}(\theta_1,\cdots,\theta_{n-1},\phi)\;,
 \end{eqnarray}
 where $Y_{l,n}$ denotes the hyperspherical harmonics on the $n$-sphere with $l$ being the
 angular momentum quantum number. By substituting the above decomposition of the scalar field into the charged Klein-Gordon equation, we can obtain the radial wave equation as following
 \begin{eqnarray}
 \frac{\Delta}{r^{n-2}}\frac{d}{dr}\left(\frac{\Delta}{r^{n-2}}\frac{dR}{dr}\right)
 +r^{2n}(\omega+eA_t)^2 R-\lambda \Delta R=0\;,
 \end{eqnarray}
 where the introduced new function $\Delta$ is explicitly given by
 \begin{eqnarray}
 \Delta=r^{2(n-2)}-\mu r^{n-1} +q^2\;,
 \end{eqnarray}
 and the separation constant $\lambda$ is given by
 \begin{eqnarray}
 \lambda=l(l+n-1)\;.
 \end{eqnarray}
 Near the event horizon $r=r_+$, the scalar field with the ingoing boundary condition
 behaves as
 \begin{eqnarray}
 \Psi\sim e^{-i\omega t} e^{-i(\omega-e\Phi_H)r_+^n r_*}\;,
 \end{eqnarray}
 where the tortoise coordinate $r_*$ is defined explicitly
 by
 \begin{eqnarray}
 \frac{dr_*}{dr}=\frac{r^{n-2}}{\Delta}\;,
 \end{eqnarray}
 and $\Phi_H=-A_t(r_+)=\sqrt{\frac{n}{2(n-1)}}\frac{q}{r_+^{n-1}}$ is the
 the electric potential at the event horizon.
 One can notice that if
 \begin{eqnarray}
 \omega<e\Phi_H\;,
 \end{eqnarray}
 the wave appears to be outgoing for an inertial observer at spatial infinity.
 This gives us the superradiant condition of scalar field in D-dimensional RN black hole.
 Since we are working with the positive
 frequency $\omega$, the superraidance will occur only for the positive charge $e$
 of the scalar field.

 In fact, we will consider the black hole in a box, i.e. the D-dimensional RN black hole
 surrounded by a reflecting mirror.
 More precisely, we will impose the mirror's boundary condition that the scalar field
 vanishes at the mirror's location $r_m$, i.e.
 \begin{eqnarray}
 \Psi(r=r_m)=0\;.
 \end{eqnarray}
 The complex frequencies satisfying the purely ingoing boundary at the black hole
 horizon and the mirror's boundary condition are called the boxed quasinormal (BQN) frequencies \cite{cardoso2004bomb}.
 The scalar modes in the superradiant regime
 will bounce back and forth between event horizon and mirror.
 Meanwhile, the energy extracted from black hole by means of
 superradiance process will grow exponentially.
 This will cause the instability of the black hole mirror system.
 In the following, we will present an analytical calculations of BQN frequencies in a certain
 limit and show the instability in the superradiant regime caused by the mirror's boundary condition.

 Now we will employ the matched asymptotic expansion method
 \cite{page, unruh} to compute the
 unstable modes of a charged scalar field in this black hole mirror system.
 We shall assume that the Compton wavelength of the scalar particles
 is muck larger than the typical size of the black hole, i.e.
 $1/\omega\gg \mu$. With this assumption, we can divide
 the space outside the event horizon into two regions, namely, a near-region,
 $r-r_+\ll 1/\omega$, and a far-region, $r-r_+\gg \mu$. The approximated solution
 can be obtained by matching the near-region solution and the far-region solution
 in the overlapping region $\mu\ll r-r_+\ll 1/\omega$. At last,
 we can impose the mirror's boundary condition to obtain the
 analytical expression of the unstable modes in this system.

 Firstly, let us focus on the near-region in the vicinity of the event horizon,
 $\omega(r-r_+)\ll 1$. It is convenient to introduce a new variable as $x=r^{n-1}$,
 the radial wave equation can be rewritten as
 \begin{eqnarray}
 (n-1)^2 \Delta\frac{d}{dx}\left(\Delta\frac{dR}{dx}\right)
 +x^{2n/(n-1)}(\omega+eA_t)^2 R\nonumber\\
 -l(l+n-1) \Delta R=0\;.
 \end{eqnarray}

 By taking the near-region limit, the radial wave function can be reduced to the form
 \begin{eqnarray}
 (n-1)^2 \Delta\frac{d}{dx}\left(\Delta\frac{dR}{dx}\right)
 +r_+^{2n}(\omega+eA_t(r_+))^2 R\nonumber\\
 -l(l+n-1)\Delta R=0\;.
 \end{eqnarray}
 Introducing another new coordinate variable
 \begin{eqnarray}
 z=\frac{x-x_+}{x-x_-}\;,
 \end{eqnarray}
 with $x_\pm=r_\pm^{n-1}$,
 the near-region radial wave equation can be rewritten in the form of
 \begin{eqnarray}
 z\partial_z(z\partial_z R(z))
 +\left[\varpi^2-\alpha\left(\alpha+1\right)\frac{z}{(1-z)^2}\right]R(z)=0\;,
 \end{eqnarray}
 where the parameter $\alpha$ and $\varphi$ are given by
 \begin{eqnarray}
 &&\alpha=\frac{l}{n-1}\;,\\
 &&\varpi=\frac{r_+^n\left(\omega-e\Phi_H\right)}{(n-1)(r_+^{n-1}-r_-^{n-1})}
 \;.
 \end{eqnarray}

 Through defining
 \begin{eqnarray}
 R=z^{i\varpi}(1-z)^{\alpha+1}F(z)\;,
 \end{eqnarray}
 the near-region radial wave equation becomes the standard hypergeometric
 equation
 \begin{eqnarray}
 z(1-z)\partial_z^2F(z)+[c-(1+a+b)]\partial_zF(z)-abF(z)=0\;,
 \end{eqnarray}
 with the parameters
 \begin{eqnarray}
 a&=&\alpha+1+2i\varpi\;,\nonumber\\
 b&=&\alpha+1\;,\nonumber\\
 c&=&\alpha+2i\varpi\;.
 \end{eqnarray}

 In the neighborhood of $z=0$, the general solution of
 the radial wave equation is then given in terms of the hypergeometric function
 \cite{handbook}
 \begin{eqnarray}
  R&=&Az^{-i\varpi}(1-z)^{\alpha+1}F(\alpha+1,\alpha+1-2i\varpi,1-2i\varpi,z)
  \nonumber\\
  &&+Bz^{i\varpi}(1-z)^{\alpha+1}F(\alpha+1,\alpha+1+2i\varpi,1+2i\varpi,z)\;.
 \end{eqnarray}
 It is obvious that the first term represents the ingoing wave
 at the horizon, while the second term represents the outgoing
 wave at the horizon. Because we are considering the classical
 superradiance process, the ingoing boundary condition at the
 horizon should be employed. Then we have to set $B=0$. The physical solution of
 the radial wave equation corresponding to the ingoing wave
 at the horizon is then given by
 \begin{eqnarray}
 R=Az^{-i\varpi}(1-z)^{\alpha+1}F(\alpha+1,\alpha+1-2i\varpi,1-2i\varpi,z)\;.
 \end{eqnarray}

 In the far-region, $r-r_+\gg M$, the effects induced by the black hole
 can be neglected. One can take the limit $\mu\rightarrow0$ and $q\rightarrow0$
 to simplify the radial wave equation in the far-region.
 In this case, we have $\Delta\simeq r^{2(n-1)}$.
 The radial wave equation reduces to the wave equation of a massless scalar field
 in the D-dimensional flat background
 \begin{eqnarray}
 \partial_r^2(r^{n/2}R(r))+\left[\omega^2-\beta(\beta+1)\frac{1}{r^2}\right](r^{n/2}R(r))=0\;,
 \end{eqnarray}
 with the parameter $\beta=l+\frac{n}{2}-1$\;.
 This equation can be solved by the Bessel function, and the general solution
 is given by \cite{handbook}
 \begin{eqnarray}
 R=r^{1/2-n/2}\left[C_1 J_{\beta+1/2}(\omega r)+C_2 Y_{\beta+1/2}(\omega r)\right]\;,
 \end{eqnarray}
 where $J_{\beta+1/2}$ and $Y_{\beta+1/2}$ are the first and the second kind
 of the Bessel functions respectively.

 In order to match the far-region solution with the near-region
 solution, we should study the large $r$ behavior of the near-region solution
 and the small $r$ behavior of the far-region solution.
 For the sake of this purpose, we can
 us the $z\rightarrow 1-z$ transformation law for the hypergeometric function \cite{handbook}
 \begin{eqnarray}
 F(a,b,c;z)&=&\frac{\Gamma(c)\Gamma(c-a-b)}{\Gamma(c-a)\Gamma(c-b)}
 F(a,b,a+b-c+1;1-z)\nonumber\\
 &&+(1-z)^{c-a-b}
 \frac{\Gamma(c)\Gamma(a+b-c)}{\Gamma(a)\Gamma(b)}\nonumber\\
 &&\times
 F(c-a,c-b,c-a-b+1;1-z)\;\;.
 \end{eqnarray}
 By employing this formula and using the properties of hypergeometric function
 $F(a,b,c,0)=1$, we can get the large $r$ behavior of the near-region solution as
 \begin{eqnarray}\label{nearsolutionlarge}
 R&\sim& A\Gamma(1-2i\varpi)\left[\frac{(r_+^{n-1}-r_-^{n-1})^{-\alpha}\Gamma(2\alpha+1)}
 {\Gamma(\alpha+1)\Gamma(\alpha+1-2i\varpi)}r^{l}\right.
 \nonumber\\&&\left.
 +\frac{(r_+^{n-1}-r_-^{n-1})^{\alpha+1}\Gamma(-2\alpha-1)}
 {\Gamma(-\alpha)\Gamma(-\alpha-2i\varpi)}r^{-l-n+1}\right]\;.
 \end{eqnarray}

 On the other hand, using the asymptotic formulas of the Bessel functions \cite{handbook},
 $J_\nu(z)=(z/2)^\nu/\Gamma(\nu+1)\;(z\ll 1)$ and $Y_\nu(z)=
 -\frac{1}{\pi}\Gamma(\nu)(z/2)^{-\nu}$, one can get the small $r$ behavior
 of the far-region solution as
 \begin{eqnarray}
 R&\sim& C_1 \frac{\left(\frac{\omega}{2}\right)^{l+\frac{n}{2}-\frac{1}{2}}}
 {\Gamma\left(l+\frac{n}{2}+\frac{1}{2}\right)}r^l\nonumber\\
 &&-\frac{C_2}{\pi}\left(\frac{\omega}{2}\right)^{-l-\frac{n}{2}+\frac{1}{2}}
 \Gamma\left(l+\frac{n}{2}-\frac{1}{2}\right)r^{-l-n+1}\;.
 \end{eqnarray}

 By comparing the large $r$ behavior of the near-region solution with
 the small $r$ behavior of the far-region solution, one can conclude
 that there exists the overlapping region $\mu\ll r-r_+\ll 1/\omega$
 where the two solutions should match. This matching yields
 the relation
 \begin{eqnarray}
 \frac{C_2}{C_1}&=&-\frac{\pi(r_+^{n-1}-r_-^{n-1})^{2\alpha+1}}{\left(l+\frac{n}{2}-\frac{1}{2}\right)
 \Gamma^2\left(l+\frac{n}{2}-\frac{1}{2}\right)}\nonumber\\
 &&\times
 \frac{\Gamma(\alpha+1)}{\Gamma(2\alpha+1)}
 \frac{\Gamma(-2\alpha-1)}{\Gamma(-\alpha)}\nonumber\\
 &&\times\frac{\Gamma(\alpha+1-2i\varpi)}{\Gamma(-\alpha-2i\varpi)}
  \left(\frac{\omega}{2}\right)^{2l+n-1}\;,
 \end{eqnarray}
 where we have used the property of Gamma function $\Gamma(x+1)=x\Gamma(x)$.

 Now we want to impose the mirror's boundary condition to study the
 unstable BQN modes. We assume that the mirror is placed near the infinity
 at a radius $r=r_m$. The far-region radial solution should vanish
 when reflected by the mirror. This yields the extra condition between
 the amplitudes $C_1$ and $C_2$ of the far-region radial solution,
 which is given by
 \begin{eqnarray}
 \frac{C_2}{C_1}=-\frac{J_{\beta+1/2}(\omega r_m)}{Y_{\beta+1/2}(\omega r_m)}\;.
 \end{eqnarray}

  This mirror condition together with the matching condition give us the following equation
 which determines the BQN frequencies of the scalar field in this black hole mirror system
 \begin{eqnarray}
 \frac{J_{\beta+1/2}(\omega r_m)}{Y_{\beta+1/2}(\omega r_m)}&=&
 \frac{\pi(r_+^{n-1}-r_-^{n-1})^{2\alpha+1}}{\left(l+\frac{n}{2}-\frac{1}{2}\right)
 \Gamma^2\left(l+\frac{n}{2}-\frac{1}{2}\right)}\nonumber\\
 &&\times
 \frac{\Gamma(\alpha+1)}{\Gamma(2\alpha+1)}
 \frac{\Gamma(-2\alpha-1)}{\Gamma(-\alpha)}\nonumber\\
 &&\times\frac{\Gamma(\alpha+1-2i\varpi)}{\Gamma(-\alpha-2i\varpi)}
  \left(\frac{\omega}{2}\right)^{2l+n-1}\;,
 \end{eqnarray}

 For the very small $\omega$, the analytical solution of BQN frequencies
 can be found from the above relation. In this case, the right hand side of the
 above relation is very small and then can be set to be zero.
 This means that
 \begin{eqnarray}
 J_{\beta+1/2}(\omega r_m)=0\;.
 \end{eqnarray}
 The real zeros of the Bessel functions were well studied. We shall label the
 $N$-th positive zero of the Bessel function $J_{\beta+1/2}$ as $j_{\beta+1/2,N}$. Then
 we can get
 \begin{eqnarray}
 \omega r_m=j_{\beta+1/2,N}\;.
 \end{eqnarray}
 In the first approximation for BQN frequencies, the solution of the Eq.(33)
 has a small imaginary part, which can be written as
 \begin{eqnarray}
 \omega_{BQN}=\frac{j_{\beta+1/2,N}}{r_m}+i\delta\;,
 \end{eqnarray}
 where the introduced imaginary part $\delta$ is small enough comparing the
 real part of BQN frequency. It can be considered as a correction to Eq.(35).
 For the small $\delta$, we can use the Taylor expansion
 of Bessel function $J_{\beta+1/2}(\omega r_m)=i\delta r_m J'_{\beta+1/2}(j_{\beta+1/2,N})$
 to proceed. Then the equation (33) can be reduced to
 \begin{eqnarray}
 i\delta r_m
 \frac{J'_{\beta+1/2}(j_{\beta+1/2,N})}{Y_{\beta+1/2}(j_{\beta+1/2,N})}
 &=&
 \frac{\pi(r_+^{n-1}-r_-^{n-1})^{2\alpha+1}}{\left(l+\frac{n}{2}-\frac{1}{2}\right)
 \Gamma^2\left(l+\frac{n}{2}-\frac{1}{2}\right)}\nonumber\\
 &&\times
 \frac{\Gamma(\alpha+1)}{\Gamma(2\alpha+1)}
 \frac{\Gamma(-2\alpha-1)}{\Gamma(-\alpha)}\nonumber\\
 &&\times\frac{\Gamma(\alpha+1-2i\varpi)}{\Gamma(-\alpha-2i\varpi)}
  \left(\frac{\omega}{2}\right)^{2l+n-1}\;,
 \end{eqnarray}

 In order to go further, we should simplify the Gamma functions on the right hand side of the
 above equation. One can note that the simplification of the Gamma function
 depends on the parameter $\alpha$. Now, we have to investigate the following two cases separately.

 \section*{Case A: $\alpha$ is an integer}

 Generally, we consider the case that the dimension $D$ of spacetime
 is greater than $4$. So we must have that $\alpha$ is an non-negative
 integer. In this case, by using the property of Gamma function $\Gamma(x+1)=x\Gamma(x)$,
 we have the following formulas
  \begin{eqnarray}
 &&\frac{\Gamma(-2\alpha-1)}{\Gamma(-\alpha)}=\frac{(-1)^{\alpha+1}\alpha!}{(2\alpha+1)!}\;,\nonumber\\
 &&\frac{\Gamma(\alpha+1-2i\varpi)}{\Gamma(-\alpha-2i\varpi)}
 =(-1)^{\alpha+1}2i\varpi \prod_{k=1}^{\alpha}(k^2+4\varpi^2)\;.
 \end{eqnarray}

 From these we can easily obtain the small imaginary part of the BQN frequencies as
 \begin{eqnarray}
 \delta=\gamma\frac{Y_{\beta+1/2}(j_{\beta+1/2,N})}{J'_{\beta+1/2}(j_{\beta+1/2,N})}
 \frac{j_{\beta+1/2,N}/r_m-e\Phi_H}{r_m^{2l+n}}\;,
  \end{eqnarray}
  where
  \begin{eqnarray}
  \gamma=\frac{\pi(r_+^{n-1}-r_-^{n-1})^{2\alpha+1}}{\left(2l+n-1\right)
 \Gamma^2\left(l+\frac{n}{2}-\frac{1}{2}\right)}
 \frac{(\alpha!)^2}{(2\alpha)!(2\alpha+1)!}\nonumber\\
 \times \prod_{k=1}^{\alpha}(k^2+4\varpi^2)\left(\frac{j_{\beta+1/2,N}}{2}\right)^{2l+n-1}\;.
  \end{eqnarray}
  One should note that the prefactor of $Y_{\beta+1/2}(j_{\beta+1/2,N})/J'_{\beta+1/2}(j_{\beta+1/2,N})$
  is always negative for our relevant range of $\omega$. Then,
  it is easy to see that, in the superradiance regime,
 $\textrm{Re}[\omega_{BQN}]-e\Phi_H<0$,
 the imaginary part of the complex BQN frequency
 $\delta>0$. The scalar field has the time dependence
 $e^{-i\omega t}=e^{-i \textrm{Re}[\omega] t}e^{\delta t}$, which implies
 the exponential amplification of superradiance modes.
 This indicates that, in this case, the scalar field with the frequency in the
 superradiant regime will undergo an instability in D-dimensional RN black hole
 surrounded by mirror.

 \section*{Case B: $\alpha$ is not an integer}

 In this case, by using the property of Gamma function $\Gamma(z)\Gamma(1-z)=\pi/\sin\pi z$,
 we have the following relations
 \begin{eqnarray}
 &&\frac{\Gamma(-2\alpha-1)}{\Gamma(-\alpha)}=-\frac{1}{2\cos\frac{\pi l}{n-1}}
 \frac{\Gamma\left(1+\frac{l}{n-1} \right)}{\Gamma\left(2+\frac{2l}{n-1} \right)}\;,\\
 &&\frac{\Gamma(\alpha+1-2i\varpi)}{\Gamma(-\alpha-2i\varpi)}
 =-\left[\frac{1}{\pi}\sin\frac{\pi l}{n-1}+
 2i\varpi\cos\frac{\pi l}{n-1}\right]\nonumber\\
 &&~~~~~~~~~~~~~~~~~~~~~~~~\times\Gamma^2\left(1+\frac{l}{n-1} \right)\;,
 \end{eqnarray}
 where we have assumed that the real part of BQN frequency
 is very near the superradiant bound and expanded the term $\Gamma(-\alpha-2i\varpi)$
 for very small $\varpi$. It should be noted that $\cos\frac{\pi l}{n-1}=0$ when
 $\frac{l}{n-1}=p+\frac{1}{2}$ with $p$ being a non-negative integer.
 In this special case, according to the discussions in Ref.\cite{mengjieprd} the asymptotic matching method fails. The numerical method must be employed to study this case.

 Here, we will just consider the case that $\frac{l}{n-1}\neq p+\frac{1}{2}$.
 Then, the real part of $\delta$ is explicitly given by
 \begin{eqnarray}
 \textrm{Re}[\delta]=\gamma'\frac{Y_{\beta+1/2}(j_{\beta+1/2,N})}{J'_{\beta+1/2}(j_{\beta+1/2,N})}
 \frac{j_{\beta+1/2,N}/r_m-e\Phi_H}{r_m^{2l+n}}\;,
 \end{eqnarray}
 where
 \begin{eqnarray}
 \gamma'&=&\frac{2(n-1)\pi}{(2l+n-1)^2}
 \frac{\Gamma^4\left(1+\frac{l}{n-1} \right)}
 {\Gamma^2\left(1+\frac{2l}{n-1} \right)\Gamma^2\left(l+\frac{n}{2}-\frac{1}{2}\right)}
 \nonumber\\
 &&\times (r_+^{n-1}-r_-^{n-1})^{1+\frac{2l}{n-1}}
 \left(\frac{j_{\beta+1/2,N}}{2}\right)^{2l+n-1}\;.
 \end{eqnarray}
 So, it is easy to see that, in the superradiance regime,
 $\textrm{Re}[\omega_{BQN}]-e\Phi_H<0$,
 the imaginary part of the complex BQN frequency
 $\delta>0$. This means that, in this case, the BQN frequencies in the superradiant regime
 is unstable for the charged scalar field in the D-dimensional RN black hole with a
 mirror placed outside of the hole.

 In summary, we have studied the instability
 of the massless charged scalar field in the D-dimensional RN black hole mirror system.
 By imposing the mirror boundary condition, we have analytically calculated
 the expression of BQN frequencies. In the first case, where $\alpha=\frac{l}{n-1}$
 is an integer, the result is very similar to the case of Kerr black hole \cite{cardoso2004bomb},
 which indicates the feature of black hole bomb.
 Especially, for the case of $n=2$, the result (39) can be reduced to the
 case of four dimensional RN black hole bomb \cite{Degolladoprd}. In the second case,
 where $\alpha$ is not an integer and $\alpha\neq p+\frac{1}{2}$ with $p$
 being an integer, the result also indicates an instability of the black hole mirror
 system. One can conjecture that, for the case of $\frac{l}{n-1}= p+\frac{1}{2}$,
 where the analytical method fails,
 the system is also unstable. However, the numerical simulation methods are
 still required to verify this conjecture in the future work.

 \section*{ACKNOWLEDGEMENT}
 This work was supported by NSFC, China (Grant No. 11205048).

 \end{document}